\documentclass[prb,showpacs,floatfix]{revtex4}
\usepackage{epsfig, amssymb}
\usepackage{color}
\usepackage{wasysym}

\begin{document}
\title{Ground-state phase diagrams of the generalized Falicov-Kimball
model\\ with Hund coupling}
\author{Romuald Lema\'{n}ski, Jakub Wrzodak}
\affiliation{Institute of Low Temperature and Structure Research, \\
Polish Academy of Sciences, 50-422 Wroc\l aw, Poland}

\begin{abstract}
Charge and spin orderings are studied on the simplest 1D and the 2D square
lattice within the generalized Falicov-Kimball model with Hund coupling
between localized and itinerant electrons. Using the restricted phase 
diagrams method (RPDM) a number of simple rules of formation of various sorts
of ground state phases have been detected. In particular, relationships
between density of current carriers (electrons or holes) and type
of charge and magnetic arrangement has been determined. In 2D in the mixed 
valence regime only axial stripes (vertical or horizontal) 
have been found for intermediate values of the coupling constants. They are 
composed of ferromagnetic or antiferromagnetic chains interchanged with 
non-magnetic ones. For band fillings close to the half filling stripe phases 
oriented along one of the main diagonal direction are formed. The results 
suggest a possibility of tuning modulations of charge and magnetic 
superstructures with a change of doping.
\end{abstract}
\pacs{71.10.Fd, 71.28.+d, 73.21.Cd, 75.10.-b}
\maketitle

\section{Introduction}
Charge and magnetic superstructures observed in many transition metal oxides,
as e.g. in $R_{2-x}Sr_xNiO_4$, where $R=La,Nd$,\cite{RKajimoto}
have stimulated an intensive search for explanation of origins of the phenomenon
and its impact on physical properties of the systems. 
The subject has been analyzed primarily in the framework of various versions of the Hubbard or \emph{t-J}
model.\cite{EmeryKivelsonLim, WhiteScalapino, PutikkaLuchiniRice, Cosentini, Oles, Kugel, Dagotto, Wrobel}

An alternative approach based on the spinless Falicov-Kimball model (FKM) 
was proposed in Ref.\cite{LemanskiFreericksBanach}.
Within the space restricted to a large number of the simplest trial configurations
the ground state diagrams have been found exactly showing how the chessboard
phase evolves to phase separation with a change of doping.
It appeared that quite large areas of the diagrams are occupied
by stripe phases oriented either along one of the main crystallographic
axes (\emph{axial stripes}), or along one of the main diagonals (\emph{diagonal stripes}).
These findings were confirmed by rigorous studies.\cite{JedrzejewskiDerzhko}

The spinless FKM is simple enough to obtain controllable
results for all values of the coupling constant. However, the model can 
deal only with charge ordered phases and it neglects magnetic properties. To overcome
this shortcoming a generalized version of the spin-one-half FKM with a spin-dependent 
local term representing Hund's first rule was proposed in Ref.\cite{RLemanski}.
An important role of the Hund coupling in explaining
magnetic properties of correlated electron systems was raised, e.g. 
in Refs.\cite{HeldVollhardt, GarciaHallbergAlascioAvignon, PruschkeBulla}, 
and specifically in applying to the FKM, in Ref.\cite{Ueltschi}.
In fact, the model we deal with is very similar to the ferromagnetic
Kondo lattice model, that was also considered in the context of charge and magnetic
superstructures in correlated electron systems\cite{GarciaHallbergAlascioAvignon, GarciaHallbergBatistaAvignonAlascio}.

Here we assume the simplest Ising-type anisotropy of the Hund coupling
what enables us to examine the model rigorously. The anisotropy 
is relevant in systems, where spin flip processes have a minor meaning and stable magnetically ordered phases occur (for more arguments see Ref.\cite{RLemanski}). 

The extended model is still oversimplified to describe all details of real materials.  But, since it comprises only basic interactions, that are present in all materials, where both localized and itinerant electrons are relevant, we expect that its characteristics emerging from our calculations are quite universal.
Our expectations are justified by the fact, that phases similar to those we detected were also found by other authors studying different models and using
quite different methods, as it was reported e.g. in Ref.\cite{Oles} for a version 
of the Hubbard model and in 
Refs.\cite{GarciaHallbergAlascioAvignon,GarciaHallbergBatistaAvignonAlascio} for the ferromagnetic Kondo (or Hund) lattice model.

In Ref.\cite{RLemanski} only some basic properties of the model in 2D were examined. 
Farka\v{s}ovk\'{y} and \u{C}en\u{c}arikov\'{a} studied the model by means of
the small-cluster exact-diagonalization calculations and an efficient numerical
method for large clusters containing up to 64 lattice sites.\cite{PFarkasovskyHCencarikova} 
They constructed phase diagrams, where they found
a number of various types of charge and spin distributions,
and observed a gradual reduction of the stability region of the non-polarized
(NP) phase in favor of the fully-polarized (FP) and partially polarized (PP)
phases with an increase of the Hund coupling and with an increase in the number 
of localized particles. The studies are interesting, as they enable to examine the model in a complementary way, but the obtained results are too general for making predictions on details of charge and spin ordering for a given set of model parameters. Besides, there are some strange irregularities in their diagrams. 
For example, for small Hund couplings or for small densities of both localized 
and itinerant particles, one can find the FP phase at some isolated 
positions, surrounded by NP phases.
And a lack of PP phases in a wide region of the diagram close
to the chessboard AF phase, especially for small $J$, seems to be an artefact
resultant from taking into considerations only clusters with even numbers of sites.

A need for a clarification of this picture pushes us to examine the model more carefully. In our previous work we considered only 2D case and we used too small configurational space to detect many regularities. We merely noticed a few general tendencies for a formation of charge and/or spin ordered phases.\cite{RLemanski}
Here we expand upon our preceding work both to 1D and 2D systems and provide a thorough
analysis of the ground state phase diagrams using a much larger
set of admissible configurations. It allows us to notice
some simple rules of formation of periodic phases (as well as their
mixtures) not noticed in previous studies.

The model Hamiltonian is
\begin{eqnarray}
\label{ham}
H= & t\sum\limits _{\left\langle i,j \right\rangle}\sum\limits _{\sigma=\uparrow ,\downarrow}
d^{\dagger}_{i,\sigma} d_{j,\sigma}
+U \sum\limits_{i}\sum\limits_{\sigma ,\eta =\uparrow,\downarrow}
n^d_{i,\sigma} n^f_{i,\eta} \nonumber \\
 & -J \sum\limits _{i}(n^d_{i,\uparrow}-n^d_{i,\downarrow} ) (n^f_{i,\uparrow}-n^f_{i,\downarrow} ),
\end{eqnarray}
where $\left\langle i,j \right\rangle $ denotes the nearest neighbor lattice sites $i$ and $j$,
$\sigma$ and $\eta$ are spin indices, $d_{i,\sigma}$  ($d^{\dagger}_{i,\sigma}$) is an annihilation
(creation) operator, and $n^d_{i,\sigma}$ ($n^f_{i,\eta}$) is
an occupation number of itinerant(localized) electrons.
The on-site interaction between localized and itinerant
electrons is represented by two coupling constants: $U$, which is
spin-independent Coulomb-type and $J$, which is spin-dependent and reflects
the Hund's rule force. The hopping amplitude $t$ is set
equal to one, so we measure all energies in units of $t$.

Double occupancy of the localized electrons is forbidden,
implying the on-site Coulomb repulsion $U_{ff}$ between two
\emph{f-electrons} is infinite.
Consequently, at a given site the \emph{f-electron} occupancy is assumed
to be $n_f=n_{f,\uparrow} + n_{f,\downarrow} \leq 1$ and the
$d-electron$ occupancy to be $n_d=n_{d,\uparrow} + n_{d,\downarrow} \leq 2$.
So there are 3 states per site allowed for the \emph{f-electrons}
($n_f=0$; $n_{f,\uparrow}=1$ and $n_{f,\downarrow}=0$;
$n_{f,\uparrow}=0$ and $n_{f,\downarrow}=1$)
and 4 states per site allowed for the \emph{d-electrons}
($n_d=0$; $n_{d,\uparrow}=1$ and $n_{d,\downarrow}=0$;
$n_{d,\uparrow}=0$ and $n_{d,\downarrow}=1$; $n_d=2$).

All single-ion interactions included in Eq. (\ref{ham}) preserve states
of localized electrons, i.e. the itinerant electrons traveling through
the lattice change neither occupation numbers nor spins of the localized
ones. Then $[H,f^+_{i\eta }f_{i\eta }]=0$ for all $i$ and
$\eta $, so the local occupation number is conserved.

The localized electrons play the role of an external, charge and spin
dependent potential for the itinerant electrons. This external potential
is "adjusted" by annealing, so the total energy of the system
attains its minimum. In other words, there is a feedback between the subsystems
of localized and itinerant electrons, and this is the feedback
that is responsible for the long-period ordered  arrangements of the localized
electrons, and consequently for the formation of various charge and/or spin
distributions at low temperatures.

In the next section we shortly describe our calculation scheme, then, 
in the third section we present two kinds of phase diagrams referring to pure 
magnets (section A) and diluted magnets (section B). The last section
contains summary and discussion.

\section{The restricted phase diagrams method (RPDM)}
We used RPDM first in our studies of the spinless 1D FKM in Ref.\cite{LachLyzwaJedrzejewski}
and then also in Ref.\cite{LemanskiFreericksBanach, RLemanski, GWatsonRLemanski}.
Within the method, calculations are performed for infinite systems 
but with a restriction to periodic phases, with periods not exceeded a certain value 
and their mixtures. Then, we can investigate both periodic phases and phase separation 
and segregation. 

We emphasize that the RPDM is by no means a mean field approach and the calculations refer to infinite systems, not to finite clusters. So we do not need to deal with neither boundary nor finite size effects. Energies (per site) of all phases we consider here are evaluated with a very high and controllable accuracy. For small period phases with no more than $4$ lattice sites in an unit cell energy bands are given by analytical expressions  \cite{RLcomment}, and the precision is limited merely by a selection of a grid in the $k-space$. For large period phases some very small errors, resulting from numerical diagonalization of matrices of size of the number of lattice sites in an unit cells may additionally enter.
The details of the current work are as follows.

We performed calculations in 1D and 2D (the square lattice) for
$U=1,2,4,6,8$ and J changing from 0.2 up to $0.75U$ and
within the configurational space restricted to all periodic phases
with unit cells containing up to 12 lattice sites for pure magnets
and up to 8 for diluted magnets.

To assure stability of the phases appearing on the diagrams,
we constructed \emph{the grand canonical phase diagrams} first
(see Ref.\cite{LachLyzwaJedrzejewski, GWatsonRLemanski} for more detailed
discussion of the stability issue) in the plane of the chemical potentials.
Then we transformed the diagrams into \emph{the canonical phase diagrams} in
the plane of densities of localized ($\rho_f$) and itinerant ($\rho_d$) electrons.
By applying this procedure one automatically includes all mixtures of the phases.
The resulting phase diagrams are quite sensitive to values of the
interaction parameters $U$ and $J$. In general, they have a rich
structure composed of various families of phases.

In order to calculate the Gibbs thermodynamic potential, we first determined
the electronic band structure for the itinerant electrons for each
candidate periodic phase. We employed a sufficiently tiny grid in the Brillouin 
zone (up to $N_c=100$ momentum points in 1D and up to $N_c=80\times80$ in 2D 
for each bandstructure).  This required us to diagonalize up to $12\times12$
matrices in the pure magnet case and up to $8\times8$ matrices in the diluted magnet case at each discrete momentum point in the Brillouin zone
and results in at most 12 and 8 different energy bands
in pure and and diluted magnet case, respectively. Hence, our calculations
can be viewed as finite size but very large cluster calculations
with cluster sizes ranging in 1D from $N=100$ up to $N=100\times12$ in the pure magnet
and from $N=100$ up to $N=100\times8$ in the diluted magnet case,
whereas in 2D from $N=80\times80$ up to $N=80\times80\times12$ in the pure magnet
and from $N=80\times80$ up to $N=80\times80\times8$
in the diluted magnet case, depending on the number of sites in the unit cell 
($N=N_cC$, where $N_c$ is equal to the number of unit cells
and $C$ denotes a number of lattice sites in unit cell for a given
configuration of localized electrons).

We performed all the calculations separately for spin up and down
itinerant electrons. The eigenvalues of the band structure are
summed up to determine the ground-state energy for each density
of the electrons. Then, the Gibbs thermodynamical potential for a given
configuration $\lbrace w _f \rbrace$
is calculated for all possible values of the chemical potentials
$\mu _d$ and $\mu _f$ of the conduction and localized electrons, 
respectively, through the formula
\begin{equation}
\label{potGibbs}
G_{\lbrace w _f \rbrace}=\frac{1}{N}
\sum\limits _{\varepsilon _\uparrow,\varepsilon _\downarrow<\mu _d}(
\varepsilon _\uparrow(\lbrace w _f \rbrace) + \varepsilon _\downarrow (\lbrace w _f \rbrace))-
\mu _d (\rho _{d\uparrow}+\rho _{d\downarrow})-
\mu _f (\rho _{f\uparrow}+\rho _{f\downarrow})
\end{equation}
where the symbol $\varepsilon _\uparrow(\lbrace w _f \rbrace)$
($\varepsilon _\downarrow(\lbrace w _f \rbrace)$) denotes
energy eigenvalues of a band structure attributed to
spin up (down) itinerant electrons for a given
configuration $\lbrace w _f \rbrace$ of localized electrons.

It appears that only a small part of the initial candidate phases can be found 
in the ground-state phase diagram. The actual number depends on $U$, $J$ and $C$
but the rate drops drastically with an increase of $C$. We find that for the values 
of the parameters we considered it is less than 10\% in 1D and less than 2\% 
in 2D case.

\section{Phase diagrams}
In this paper we present two types of the ground-state phase diagrams. The first type (\emph{pure magnets}) demonstrates only magnetic order, as it corresponds to the case $\rho_f=1$ (each site is occupied by exactly one \emph{f-electron}) in the plane $(J,\rho_d)$. And the second type (\emph{diluted magnets}) demonstrates both
a magnetic and charge order in the plane $(\rho_d,\rho_f)$ for fixed values $J$ and $U$. The diagrams show ground state configurations of the \emph{f-electrons} both in 1D and 2D for representative values of the model parameters. For a pure magnet we selected $U=4$, $0.2\leq J\leq 3.0$ in 1D and $U=6$, $0.2\leq J\leq 3.0$ in 2D. And for the diluted magnet $U=2$, $J=0.5$ in 1D and  $U=4$, $J=0.5$ in the 2D.

\subsection{pure magnets}
In the pure magnetic diagrams the ferromagnetic phase (F) is stable for 
$\rho_d$ close to 0 or 2 and the region of the stability increases 
with $J$, whereas along the line  $\rho_d=1$  (the half-filling) 
the simplest AF phase is stable. Now, the most interesting story concerns 
a way of transforming between the two extreme phases with a change 
of $\rho_d$.

\begin{figure}[ht]
    \epsfxsize=12cm
    \epsffile{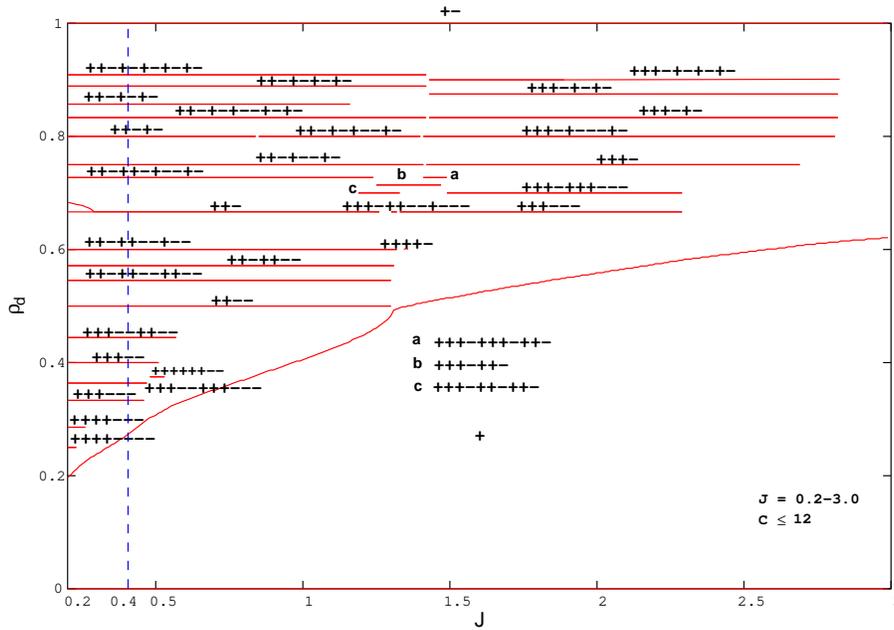}
\caption{The one-dimensional pure magnetic phase diagram restricted to all periodic 
phases with $\rho_f=1$ and with the maximum period $C\leq 12$. Straight line segments mark stability intervals of the phases. Unit cells of the phases are expressed
by sequences of the plus and minus signs placed close to (in almost all cases
just above) the corresponding line segments. The signs ``+'' and ``-''denote
up and down spins of the \emph{f-electrons}, respectively. The extended area
below the curve line at the bottom of the diagram shows a stability region
of the ferromagnetic phase F. Unit cells of phases located along the dashed vertical line for $J=0.4$ are displayed in Table I.}
\label{fig1}
\end{figure}
\begin{figure}[ht]
    \epsfxsize=12cm
    \epsffile{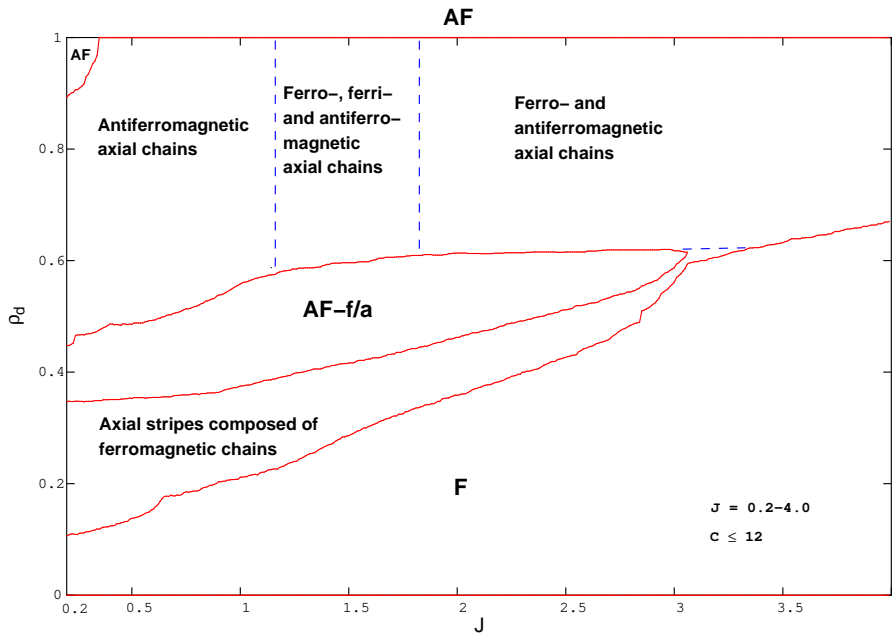}
\caption{The two-dimensional pure magnetic phase diagram restricted to all periodic phases with $\rho_f=1$ and $C\leq 12$. Typical configurations of spins of the  \emph{f-electrons} representing phases from particular regions of the diagram are shown in Figs. 3. }
\label{fig2}
\end{figure}
\begin{figure}[ht]
    \epsfxsize=12cm
    \epsffile{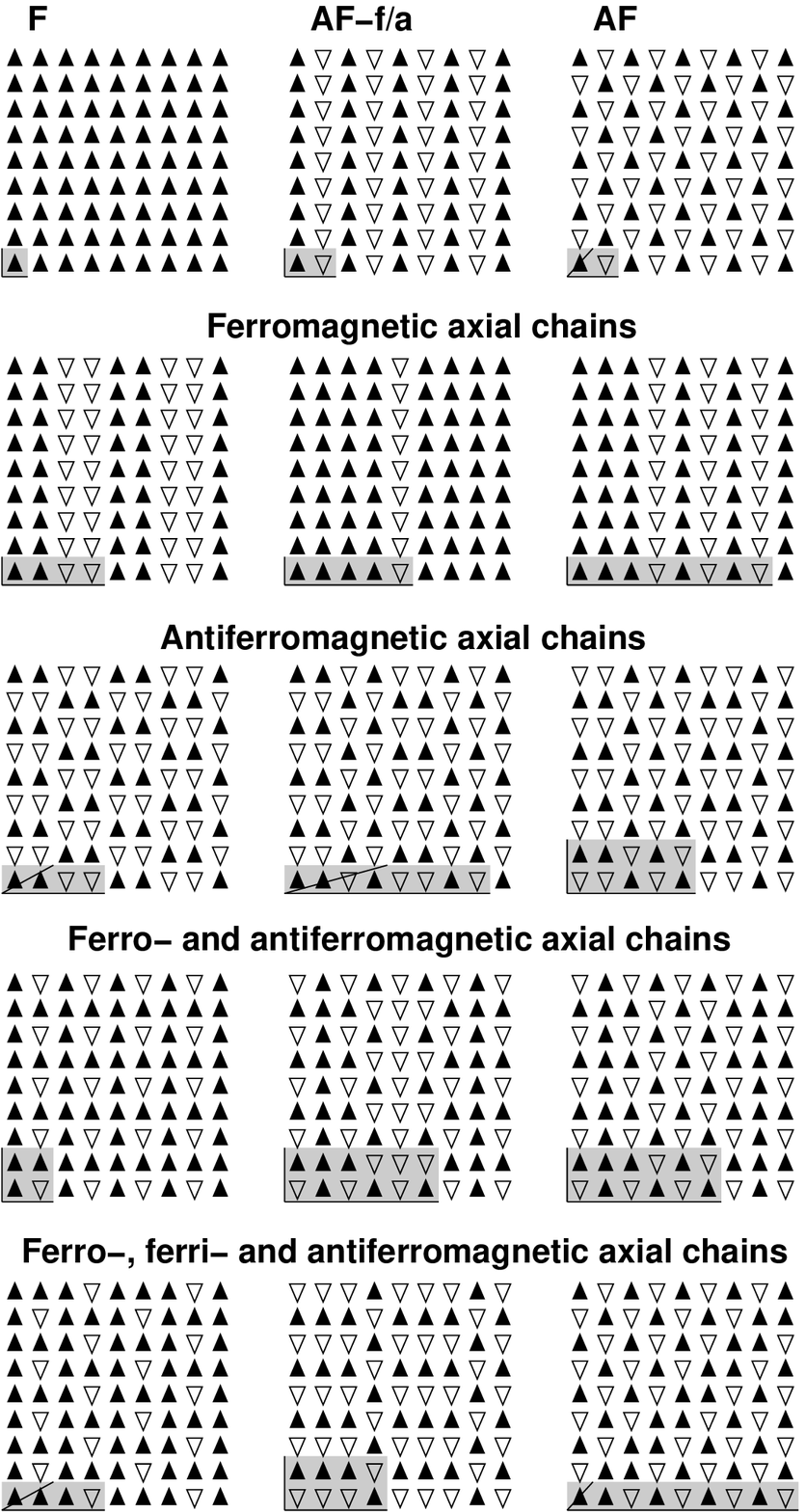}
\caption{Examples of ground-state periodic phases found in th the diagram displayed
in Fig. 2. The symbol $\blacktriangle$ ($\triangledown $)
denotes a spin up(down) \emph{f-electron}. The shaded rectangulars
in the left bottom parts of the pictures mark unit cells of the corresponding 
phases and the straight line segments mark the translation vectors. }
\label{fig3}
\end{figure}

Obviously, the process depends on $J$, but it is the density $\rho_d$ that plays a crucial role in determining a spin order. Namely, if $\rho_d=p/q$, where $p$ and $q$ are relative prime numbers, then the period $r$ of a stable phase in 1D is equal to $q$ or a multiple of $q$ (i.e. $r=nq$, $n=1,2,...$). Consequently, if $r=q$ and $q$ is an odd number, then the system cannot be an antiferromagnetic (AF), but ferrimagnetic (FI).
Indeed, we observe both FI and AF phases distributed over the whole region between the F and the simplest AF phases. This is in contrast to the results reported in Ref.\cite{PFarkasovskyHCencarikova}, where many FI phases (named as partially polarized PP) were missed in 1D because only systems containing even numbers of lattice sites
were taken into account. On the other hand, our AF phases are consistent with NP (non polarized) phases reported in Ref.\cite{PFarkasovskyHCencarikova}.

It appears that not only the period is determined by $\rho_d$.
We found a remarkable feature concerning the number $L_f$ of changes
of the \emph{f-electron} spin orientation (from up to down or from down
to up) calculated per site. If in the diagram displayed in Fig. 1 we move up 
along a vertical line (i.e. when $J$ is fixed) then $L_f$ of subsequent phases
increases with the density $\rho_d$. What is more, for $J\leq 1$
in almost all cases $L_f=\rho_d$. Then the number of itinerant
electrons is equal to the number of pairs of localized electrons 
with magnetic moments oriented oppositely.

Physically this rule means that each moving electron is somehow associated with 
an exactly one abrupt change of the potential resulting from the localized electrons.
In other words, the minimum energy is attained when the number of moving electrons
and the number of changes of the potential acting on them are equal to each other.
The rule can be noticed by direct inspection, e.g. looking along the dashed line in Fig. 1 
(for $J=0.4$). In this case unit cells of phases located between F and the simplest 
AF phases are displayed in Table I.

\begin{table}[h]
\begin{tabular}{||cr|cr||cr|cr||cr|cr||}
\hline
unit cell&&$\rho_ d$&&unit cell&&$\rho_ d$&&unit cell&&$\rho_ d$& \\
\hline
$+++---$&&1/3&&$+++--+++---$&&4/11&&$+++--$&&2/5& \\
\hline
$+++--++--$&&4/9&&$++--$&&1/2&&$++-++--++--$&&6/11& \\
\hline
$++-++--$&&4/7&&$++-++--+--$&&3/5&&$++-$&&2/3& \\
\hline
$++-++-+--+-$&&8/11&&$++-+--+-$&&3/4&&$++-+-$&&4/5& \\
\hline
$++-+-+--+-+-$&&5/6&&$++-+-+-$&&6/7&&$++-+-+-+-$&&8/9& \\
\hline
$++-+-+-+-+-$&&10/11&&$+-$&&1&& && & \\
\hline
\end{tabular}  
\caption{Unit cells of phases located along the dashed line $J=0.4$
in Fig. 1 and electron densities $\rho_ d(=L_f)$ corresponding to them.}
\end{table}
Obviously, for small enough $\rho_d$, where the F phase is stable,
one has $L_f=0$ and for $\rho_d=1$, where the simpest AF phase
is stable, one has $L_f=1$.
So it is clear that in 1D the density of itinerant electrons $\rho_d$
determines not only a periodicity (within an accuracy to a small natural
number multiplier) of arrangement of the \emph{f-electrons} but also
strongly influences a relative distribution of spins up and down inside
unit cells.

In 2D the process of transformation from F to AF with an increase of $\rho_d$ can be divided into two stages (see Fig. 2). First, anisotropic quasi one-dimensional structures composed of parallel ferromagnetic chains oriented along one of the main lattice axis are formed. We call the area 
\emph{the region of axial stripes with ferromagnetic chains} (see Figs. 2 and 3). For $J\apprle 3.05$ this region ends up with the simplest phase belonging to this class, that is composed of ferromagnetic chains with alternating spin direction. In our considerations this is the very special phase, as it can be also viewed as composed of the simplest antiferromagnetic chains along the perpendicular axis. This is why we call the phase AF-f/a, to underline that it is the antiferromagnetic phase composed of ferro-/antiferro- magnetic chains (see Figs. 2 and 3).

Above the stability region of AF-f/a a majority of phases (see Fig. 3) 
are composed of either only the simplest antiferromagnetic chains 
(for $J\apprle 1.2$) or with an admixture of ferromagnetic chains 
(for $J\apprge 1.8$) and in the intermediate interval 
of $1.2\apprle J\apprle 1.8$ also of ferrimagnetic chains. 
Some phases found in this region can be viewed as composed of diagonal 
ferromagnetic chains oriented along the diagonal (1,1) direction.
And the final stage of the transformation of the phases
with an increase of $\rho_d$ is the simplest AF phase with
antiferromagnetic chains located along the both main lattice axes.

It appears that the transformation from F to the simplest AF phase is
accompanied with an increase of a rate of localization of itinerant
electrons, as with an increase of $\rho_d$ a mobility 
of the \emph{d-electrons} becomes more
and more restricted when the half-filling is approached.
For small $\rho_d$, where the F phase
is stable, the \emph{f-electrons} act on the \emph{d-electrons}
as an uniform, site independent external field that don't disturb
their movements. Then, in the region
of \emph{axial stripes with ferromagnetic chains} the \emph{d-electrons}
can move easily but only along these chains, as along the perpendicular 
direction an external potential (coming from the \emph{f-electrons}) 
alternates by taking two different values $U+J$ and $U-J$
what causes scattering of the \emph{d-electrons}.

The AF-f/a phase is an optimal one with respect to the transport of the
\emph{d-electrons} through the lattice but only along one direction.
Maybe it is related to the optimum doping reported
in some materials, when there is a balance between a density
of current carriers and their mobility over the lattice.
Obviously, one should be cautious when trying to relate the results
obtained for such a simple model with situations observed in
real materials, but it is interesting that the optimum doping
observed here is attained for $\rho_d$ close to 0.5, what corresponds
to the special case of quarter filling.

A further increase of $\rho_d$ causes a complete vanish of ferromagnetic chains
for small values of $J$ ($J\apprle 1.2$) and a gradual decrease of their number
for not too small $J$.
It means that the \emph{d-electrons} meet more and more potential
barriers in any direction what makes them more and more localized.
Obviously, the rate of localization becomes higher when $J$ is large.

Here we point out another interesting feature of the model. Namely, the critical value $\rho_d^*$ below which phases containing antiferromagnetic chains are stable increases with $J$, so the range of densities $\rho_d$ where the \emph{d-electrons} become more
localized shrinks, but at the same time a rate of the localization becomes more pronounced. It means that for large $J$ the \emph{d-electrons} pass from a delocalized to localized regime within a relatively small interval of their densities. And the reported results suggest that in the limit of infinite $J$ the interval between the conducting F and insulating AF phase tends to zero (close to the line $\rho_d=1$). This may be regarded as an analogy to  the famous Nagaoka problem studied within the Hubbard model.

\subsection{diluted magnets}
Let us now analyse a phase diagram corresponding to a diluted
magnet, where both spin and charge orderings are relevant.
The diagram is displayed in the ($\rho_f,\rho_d$) plane
for $U=2$ and $J=0.5$ in the 1D case (see Fig.4) and for
$U=4$ and $J=0.5$ in the 2D case (see Fig. 5).
The maximum period C of allowed phases in the two cases
is equal to 8. The values of the parameters $U$ and $J$
were chosen to be characteristic intermediate value representatives.

\begin{figure}[ht]
    \epsfxsize=12cm
    \epsffile{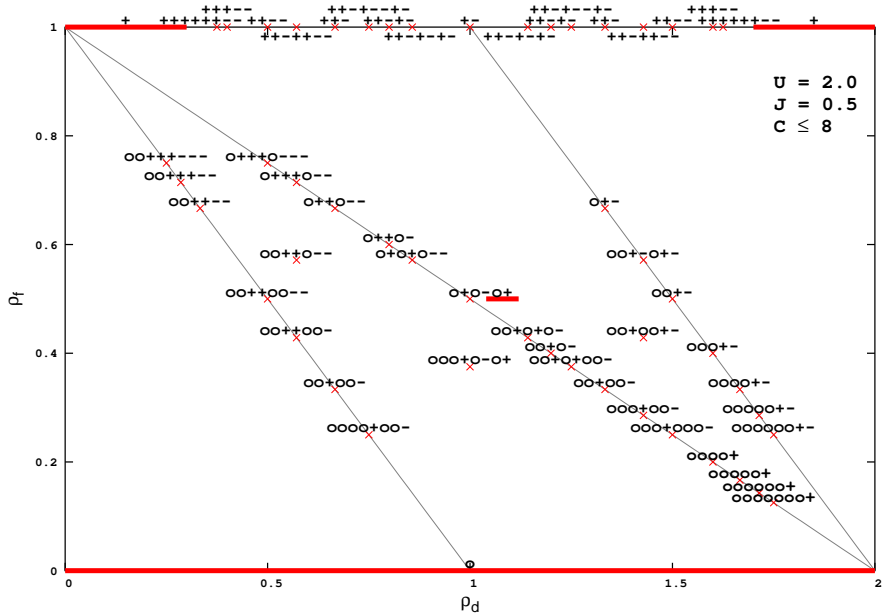}
\caption{The canonical phase diagram of the extended FKM with Hund coupling
for the 1D lattice and $U=2$, $J=0.5$.
The red crosses \color{red}$\times$\ \color{black} and horizontal straight line 
segments mark stability points or intervals of periodic phases. Their unit cells 
are drawn as sequences of small circles and plus and minus signs that correspond
to sites non-occupied, occupied by the spin up
and occupied by the spin down \emph{f-electrons}, respectively.}
\label{fig4}
\end{figure}
\begin{figure}[ht]
    \epsfxsize=12cm
    \epsffile{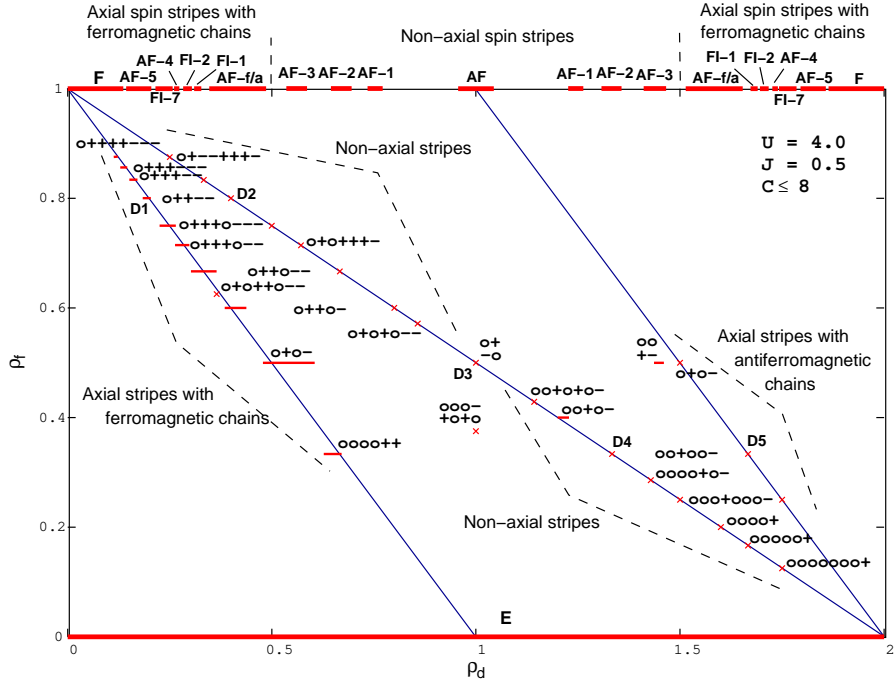}
\caption{The canonical phase diagram of the extended FKM with Hund coupling
for the 2D square lattice and $U=4$, $J=0.5$.
The lines $\rho_f=1-\rho_d$,
$\rho_f=2-\rho_d$ and $\rho_f=1-\rho_d/2$ are merely visual guides. The red crosses 
\color{red}$\times$\ \color{black} and horizontal straight line segments mark 
stability points and intervals of the periodic phases, respectively. Their unit 
cells are drawn as sequences of small circles and plus and minus signs that
correspond to sites non-occupied, occupied by the spin up
and occupied by the spin down \emph{f-electrons}, respectively.
A number of pairs of phases have the same unit cells but different
translation vectors. Unit cells of the phases are displayed
along the horizontal lines in the middle
between the lines $\rho_f=1-\rho_d$ and $\rho_f=1-\rho_d/2$,
and in the middle between
the lines $\rho_f=1-\rho_d/2$ and $\rho_f=2-\rho_d$.
The configurations located along the line $\rho_f=1$ are presented
in Fig. 3 and a set of characteristic configurations D1, D2, D3, D4, D5
is shown in Fig. 6.}
\label{fig5}
\end{figure}
\begin{figure}[ht]
    \epsfxsize=12cm
    \epsffile{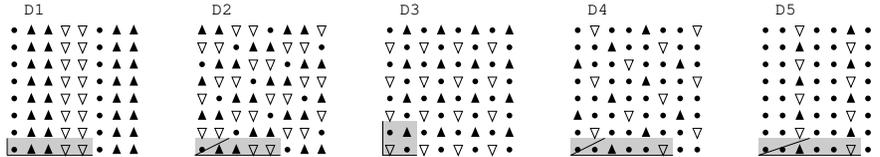}
\caption{Characteristic ground state configurations displayed
in Fig. 5. See the caption to Fig. 3 for more explanations.}
\label{fig6}
\end{figure}

We found both in the 1D and 2D diagrams that a majority of periodic phases 
are located along one of the following three lines: 
$\rho_f=1-\rho_d$, $\rho_f=2-\rho_d$ or the diagonal $\rho_f=1-\rho_d/2$. 
The first two mentioned lines correspond to mixed valence regimes.

Both antiferro- and ferrimagnetic arrangements of the
\emph{f-electrons} are found in the whole range of $\rho_f$ and
$\rho_d$ in 1D and 2D. In 1D unit cells
of phases located along the line $\rho_f=1-\rho_d$ are composed
of blocks of spins up (+) and down $(-)$,
whereas the pairs of opposite spins $(+-)$ are stable along
the $\rho_f=2-\rho_d$ line. Unit cells of phases located
along the diagonal $\rho_f=1-\rho_d/2$ have the most homogeneous types
of structures. A typical example of the transformation can be noticed
e.g. for $\rho_f=2/3$, where the unit cell $\lbrace oo++--\rbrace$
transforms first to $\lbrace o++o--\rbrace$ and then to
$\lbrace o+-\rbrace$ for $\rho_d=1/3, 2/3$ and $4/3$, respectively.

In 2D (see Figs. 5 and 6), phases located along the
$\rho_f=1-\rho_d$ line are composed of ferromagnetic
(or diluted ferromagnetic) and nonmagnetic chains oriented along
one of the lattice axis (e.g. D1 in Fig. 6).
Phases belonging to this family are marked on the diagram in Fig. 5
by straight line segments. It means that they are stable over
finite intervals of band fillings.

On the other hand, phases located along the $\rho_f=2-\rho_d$
line are composed of antiferromagnetic and nonmagnetic chains.
And phases located along the diagonal $\rho_f=1-\rho_d/2$
can be viewed as composed of diluted ferro- or 
antiferromagnetic chains (D2, D3 and D4 in Fig. 6).
The highest symmetry has the phase D3 placed at the central
point of the diagram ($\rho_f=1/2$, $\rho_d=1$).

It is interesting, that phases located along the diagonal
$\rho_f=1-\rho_d/2$ are insulating for any values of the model 
parameters we examined, as they have gaps at their
Fermi levels, whereas phases found along the line
$\rho_f=1-\rho_d$ have no energy gaps at their Fermi levels.
And phases located along the line $\rho_f=2-\rho_d$ have no gaps 
for small values of $U$, but they do have gaps for $U$ large enough.
This is consistent with the conjecture that the \emph{d-electrons} 
can easily (i.e. without scattering) move along ferromagnetic and nonmagnetic chains, 
but along antiferromagnetic chains their mobility becomes suppressed

\section{Summary and discussion}
Since the diagrams reported in this paper were constructed within the restricted 
space of periodic configurations, they can serve only as skeletons of the full diagrams. Here, similarly to what was found in the case of the simplest FKM,\cite{LemanskiFreericksBanach} most of the diagrams areas are occupied by mixtures of various phases, occasionally penetrated by periodic phases.

With an increase of the maximum period $C$ of admissible configurations
more and more periodic phases with higher periods replace some of the mixtures 
on the canonical phase diagrams. However, we observed that the higher period phases do not distroy the diagrams' structure, i.e. charge and spin distributions of these new  phases follow the same rules that we already detected for low period phases. So our conjecture is that the full diagrams will be filled with phases of which charge and magnetic order can be easily predicted (for a given set of the coupling parmeters and densities $\rho_ d$ and $\rho_ f$). Of course, working within the RPDM we are not able to prove the statement rigorously, but since it appears to be quite reasonable, we expect that it can be established definitely by other methods.

In the limiting case of $C$ tending to infinity not only periodic,
but also aperiodic phases may happen to appear on the diagrams. 
It is not clear if some mixtures of low period phases survive in the central 
region of the full phases diagram. But it is quite possible, as in the simplest 
spinless FKM such phases are proven to have the lowest energy in the large $U$ limit 
\cite{TKennedy}. 

The rules of formation of the phases we detected from an analysis 
of the diagrams do not allow to determine unambiguously the ground state charge 
and spin arrangement for given values of $\rho_ f$, $\rho_ d$, $U$ and $J$,
but they provide enough information needed for a rough prediction 
of what sorts of phases appear on the digrams and where they are located.

In the pure magnetic case ($\rho_ f=1$) the F phase is stable for the densities
$\rho_ d$ such that $\rho_ d < \rho_d^*(J)$ or $2-\rho_d^*(J)<\rho_ d$, where
$\rho_d^*(J)$ is an increasing function of $J$. Within the interval of $J$
ranging from 0.2 to 3.0 the function $\rho_d^*(J)$ increases from about 0.2
to slightly above 0.6 in 1D (see Fig. 1) and from about 0.1 to around
0.55 in 2D (see Fig.2). The results are consistent with the data obtained
in Ref.\cite{PFarkasovskyHCencarikova}.

When $\rho_ d$ tends to the half filling $\rho_ d=1$, a transformation 
from F to the simplest AF phase occurs in 1D according to the following 
simple rules.

1. If $\rho_ d=p/q$, where $p$ and $q$ are relative prime integers,
then if a phase is periodic, then its period is equal to $nq$ ($n=1,2,...$).

2. For $J$ small and $\rho_ d=p/q$, with $q$ being an even integer, periodic 
phases are antiferromagnetic, whereas for $q$ being an odd number they are 
ferrimagnetic with the lowest possible magnetization; for large $J$  
higher magnetizations states become stable.

3. For a given $J$ the number $L_f$ of changes of spin orientation
calculated per site increases with $\rho_ d$ and for small $J$ it is equal
to $\rho_ d$.

4. For a given density $\rho _d$ the number $L_f$ drops with an increase
of $J$.

The rules confirm a presence of quite well organized phase diagram
structure not revealed in previous studies. In fact, some of the details shown
in Ref.\cite{PFarkasovskyHCencarikova} as, for example, arrangements of spins
in a certain number of phases are in agreement with these rules. However, since
only rings composed of even numbers of sites and even numbers of electrons were 
investigated in,\cite{PFarkasovskyHCencarikova} a number of FI phases were missed.

Driving mechanisms that are behind the detected rules are still not fully understood. 
Recently Brydon and Gulacsi \cite{PMRBrydonMGulacsi} discovered that competitive roles of the forward-scattering and back-scattering of itinerant electrons  can explain observed richness of the spinless FKM diagrams.
We hope that studies carried out along the similar ways could be also performed for the extended version of the FKM with the Hund coupling and elucidate the rules we observed.

In 2D the situation is more complex and we were not able to find out as many precise rules as in the 1D case. Even though, our phase diagram shows more regularities than those of reported in Ref.\cite{PFarkasovskyHCencarikova}. First of all, we noticed that all phases appeared in the diagram are composed of ferromagnetic or antiferromagnetic, 
and for intermediate values of $J$ also ferrimagnetic chains parallel to each other. Obviously, the phases with only ferromagnetic chains have one-dimensional unit cells and they form axial stripes. These phases occur within an interval of electron densities $\rho _d$ neighboring to those for which the F phase is stable. For $J\apprle 3.05$ the interval ends with the simplest phase belonging to the family, the AF-f/a phase (see Fig. 2), that separates regions of axial stripes from those of containing antiferromagnetic chains (ferromagnetic and ferrimagnetic chains could be also present for not too small values of $J$). So for $\rho _d$ out of the stability regions of F, AF-f/a and axial stripes almost all phases are composed of either exclusively antiferromagnetic chains or with an admixture of ferri- and ferromagnetic chains. Some of them containing only ferro- and antiferromagnetic chains are ferrimagnetic.

\vspace{0.3cm}
An analysis of diluted magnets diagrams (see Figs. 4-6) 
also permits us to fix some rules of charge and spin formation and its 
evolution with a change of the densities $\rho _d$ and $\rho _f$. 
Here we focused on  the most representative three 
families of the phases. One of them consists of
phases located along the main diagonal. This family corresponds 
to the most homogeneous phases relevant for the spinless FKM, and this
is the only family of diluted periodic phases which is left in the limit 
of large $U$ (if we keep $J$ considerable smaller than $U$).
Phases belonging to this family are characterized by the most uniform
charge distribution but not necessarily the most uniform magnetic 
distribution. In 2D all but one particular phase have a form of sloped 
stripes composed of parallel lines of ferromagnetic chains 
(see conf. D2 and D4 in Fig. 6).

The only exception is the most symmetric, antiferromagnetic chessboard 
phase D3 placed in the center of the diagram. The phase has two-dimensional
unit cell of size $2\times 2$ and is composed of diluted ferromagnetic 
lines (see Fig. 6).

Two other characteristic families refer to mixed valence regimes,
for which either the condition $\rho _d+\rho _f=1$ or $\rho _d+\rho _f=2$
is fulfilled. These phases are ground states only for small and
intermediate values of $U$ (and $U>>J$). In 1D, it appears that 
unit cells of phases belonging the first category ($\rho _d+\rho _f=1$) 
are built of blocks of spins up separated by pairs of empty sites from 
blocks of spins down. On the other hand, unit cells of phases
belonging to the second category ($\rho _d+\rho _f=2$) consist of empty
sites separated by pairs of oppositely oriented spins ($+-$). 

In 2D, all phases coming from the both mixed valence categories have 
the form of axial stripes. So they have the same type of charge ordering.
Nevertheless their magnetic orders are clearly different, 
as phases that belong to the first class
are composed of ferromagnetic chains (e.g. D1 in Fig. 6), whereas
phases for which the condition $\rho _d+\rho _f=2$ is fulfilled
are composed of antiferromagnetic chains (e.g. D5 in Fig. 6).

Our current studies confirm findings reported in
Ref.\cite{Oles} that show that the compromise between kinetic energy
of the \emph{d-electrons} and their interaction with the \emph{f-electrons}
imposes formation of superstructures with shapes of stripes.
Kinetic energy tends to spread out the \emph{d-electrons} uniformly 
over the lattice, but due to the presence of localized
magnetic ions a kind of \emph{d-electron} density deformation 
must occur. Obviously, the deformation has to be 
conjugated with an arrangement of the \emph{f-electrons}. 
Apparently, the simplest departures from the homogeneity that are
preferred have the form of axial or diagonal stripes.

Perhaps the most important conclusion emerging from this work
is that the observed rules of formation of the phases 
suggest a possibility of manipulation of positional arrangements 
of magnetic ions diluted in the system and also their magnetic alignment
with a change of doping. For example, one should be able to tune
a modulation of charge and/or spin (stripes' width).
If it can be done in a controllable way,
then in systems that can be described by the model it would be possible to
change gradually an orientation of stripe phases (between axial 
and diagonal) and to change magnetic order along chains 
(from ferro- through ferri- up to antiferromagnetic).

We hope the results will motivate some new experimental work focusing on
searching relationships between density of current carriers
(electrons or holes) and observed charge and/or magnetic superstructure.
According to our findings complicated ordering patterns should
emerge from on-site interactions of localized and moving electrons
and a simplified version of Hund's rule. Therefore, we expect that
experimental realizations of such patterns are robust in those correlated
electron systems where a substantial anisotropy of spin-spin interactions occur.

\acknowledgements
We thank J.K. Freericks, M. Ma\'ska and J. J\c{e}drzejewski for valuable comments and critical reading of the manuscript.


\begin{references}
\bibitem{RLcomment}
	Only these kind of phases were taken into account in our previous papers where the model was considered.
\end{references}
\end{document}